\documentclass[11pt,fleqn]{article}
\newif\ifpdf
\ifx\pdfoutput\undefined
    \pdffalse    % we are not running pdfLaTeX
\else
    \pdfoutput=1 % we are running pdfLaTeX
    \pdftrue
\fi

\ifpdf
   \usepackage[pdftex,final]{graphics}
\else
   \usepackage[dvips,final]{graphics}
\fi

\usepackage{fancyhdr,ifthen,amssymb,here}
\usepackage[pdftex]{graphicx}
\usepackage{epsfig}

\newlength{\textwidthMinusEight}
\newfont{\bigrm}{cmr10 at 12pt}
\newfont{\Bigrm}{cmr12 at 16pt}

\newcommand{\la}{\lambda}

\newcommand{\mr}{\mathrm}
\newcommand{\mb}{\mathbf}
\newcommand{\nn}{\nonumber\\[1.5ex]}

\def\fs{\; \; .}
\def\co{\; \; ,}

\newcommand{\mpi}{M_{\pi}}

\newcommand{\mpiL}{M_{\pi L}}
\newcommand{\lapi}{\lambda_{\pi}}

\newcommand{\fpi}{F_{\pi}}

\newcommand{\lagr}{\mathcal{L}}
\newcommand{\lb}{\bar\ell}

\newcommand{\unith}{{1\hspace*{-1.5mm}1}}
\newcommand{\<}{\langle}
\renewcommand{\>}{\rangle}

\newcommand{\til}{\tilde}

\newcommand{\cO}{\mathcal{O}}
\newcommand{\cF}{\mathcal{F}}
\newcommand{\cP}{\mathcal{P}}
\newcommand{\cH}{\mathcal{H}}
\newcommand{\ri}{\mr{i}}

\begin{document}
\DeclareGraphicsExtensions{.pdf,.png,.jpg}

\vspace*{0.8cm}
\begin{center}
{\Bigrm Finite volume effects for the pion mass at two loops}
\vskip 0.6cm
{\bigrm Gilberto Colangelo and Christoph Haefeli}
\vskip 3ex
{\bigrm\it Institut f{\"u}r Theoretische Physik, Universit{\"a}t Bern}
\\
{\bigrm\it Sidlerstr. 5, 3012 Bern,  Switzerland } 
\end{center}

\vskip 0.8cm
\hrule
\vskip 2.0ex
{\small
\setlength{\baselineskip}{14pt}
\noindent
{\bf Abstract}
\vskip 1.0ex\noindent
We evaluate the pion mass in finite volume to two loops within Chiral
Perturbation Theory. The results are compared with a recently proposed
extension of the asymptotic formula of L\"uscher. We find that contributions,
which were neglected in the latter, are numerically very 
small at the two--loop level and conclude that for $\mpi L\gtrsim 2$,
$L\geq2\mr{fm}$ the finite volume effects in the meson sector are
analytically well under control. }

\vskip 2.0ex
\hrule
\vspace*{0.5cm}

\setlength{\baselineskip}{15pt}

\section{Introduction}

\noindent
Numerical simulations in lattice QCD are bound to rather small volumes.
When one determines the hadron spectrum and other low energy observables
one has to understand and properly account for the volume dependence in
order to correctly interpret the numerical data. Analytical methods which
allow one to predict the size of the finite volume effects are particularly
useful in this respect. In the case of hadron masses, there are two
different methods to analytically evaluate the finite volume effects: the
asymptotic formula derived by L\"uscher~\cite{Luscher:1985dn}, and chiral
perturbation theory (ChPT) in finite
volume~\cite{Gasser:1986vb,Gasser:1987ah,Gasser:1987zq}. L\"uscher's
formula relates the volume dependence of the mass of a hadron to an
integral over the $\pi$-hadron forward scattering amplitude. Knowledge of
the latter scattering amplitude immediately translates into an estimate of
finite volume effects for the hadron mass. Since the integral is dominated
by the low-energy region, one can rely on the chiral representation of the
scattering amplitude in order to numerically evaluate the integral. In this
manner one makes use of ChPT only in infinite volume and obtains an
estimate only of the leading exponential term in the finite volume
dependence, the term of the order $\exp(-M_\pi L)$, where $L$ is the box
size. Alternatively one can perform the calculation of the hadron mass in
ChPT in finite volume and obtains an estimate also of the terms which are
exponentially subleading.

A detailed study and comparison of the two approaches has been performed in
\cite{Colangelo:2002hy,Colangelo:2003hf} for the case of the pion mass.
This case is particularly interesting both because it is a simple hadron to
be studied on the lattice and because the $\pi \pi$ scattering amplitude,
which is needed in the L\"uscher formula, is now known to
next-to-next-to-leading
order~\cite{Gasser:1983yg,Bijnens:1995yn,Bijnens:1997vq,Colangelo:2001df}
in the chiral expansion. It was therefore possible to study numerically how
well the leading exponential term dominates the series, and how fast the
chiral expansion (of the leading exponential term) converges. The somewhat
surprising result was that the leading term in both series receives large
corrections both from subleading exponentials as well as from the
next-to-leading order in the chiral expansion. The next-to-next-to-leading
order chiral correction, on the other hand, was found to be rather small,
which indicates that -- at this level -- the chiral series has started
converging well.

The fact that the L\"uscher formula appeared to give the numerically
dominating term only for volumes so large that the finite volume correction
itself has become negligible appeared to be more worrisome. At first sight
this might have lead one to conclude that in cases of practical interest
one could not rely on this extremely convenient formula in order to
evaluate reliably the finite volume effects. We believe, however, that such
a negative conclusion is unjustified and have proposed a resummation of the
L\"uscher formula~\cite{Colangelo:2004sc,Colangelo:2005gd} which retains
all the convenience of the original one but does not suffer from the same
large corrections. This resummation can be understood in simple terms:
L\"uscher has shown that the leading exponential correction comes from a
radiative correction in which the emitted virtual light particle (the pion,
in QCD) goes around the world once (thanks to the periodic boundary
conditions) before being reabsorbed. The resummed formula takes into
account all other possible ways which the pion has to go around the world
(go along the diagonal, or go around more than once, etc.).

As was shown in~\cite{Colangelo:2005gd} this resummed formula exactly
reproduces the one--loop ChPT calculation of the pion mass in finite volume
if one inserts in the integral the leading chiral representation of the
$\pi \pi$ scattering amplitude. Inserting the next-to-leading chiral
representation of the scattering amplitude one reproduces the full two--loop
calculation of the pion mass in ChPT in finite volume up to corrections of
order $\exp(-\bar M L)$ with $\bar M \geq (\sqrt{3}+1)/\sqrt{2} M_\pi$.
Despite this improvement of the algebraic accuracy of the resummed formula
with respect to the L\"uscher's one, it is essential to show that the
resummation is numerically effective and that the corrections to it are
small. In order to make this check we have now performed a full two loop
calculation of the pion mass in finite volume in ChPT. As we will
demonstrate in this article, the corrections which are not captured by the
resummed asymptotic formula are negligibly small for $\mpi L \gtrsim 2$.
The results of the present two--loop calculation were anticipated in
\cite{Haefeli:2005px,Colangelo:2005cg}. Here, we give further details about
the calculation and the results.
 
One of the reasons for performing this further investigation of finite
volume effects for hadron masses is that one may view L\"uscher's formula
(or its resummed version) as a way to determine on the lattice a scattering
amplitude, cf.~\cite{Colangelo:2005gd}. While it is true that the
scattering amplitude is seen here in an exponentially suppressed effect, it
is also true that the direct calculation of a scattering amplitude on the
lattice is much more difficult than that of a mass. This indirect method to
extract a scattering amplitude may in some cases turn out to be more
practical, and we find it worthwhile to discuss its theoretical
feasibility. Indeed if the part of the finite volume corrections which is
related to the scattering amplitude is not strongly dominating with respect
to the rest, this is not a viable method to determine the scattering
amplitude. The conclusion reached in this paper is therefore encouraging in
this respect and can be used to estimate in which region of the $(M_\pi,L)$
plane the finite volume correction to a hadron mass is given to a good
approximation by the integral containing the scattering amplitude.

The two--loop calculation is interesting in its own right, also from a
technical viewpoint. To date, a number of finite volume calculations have
been performed at one--loop order
\cite{Sharpe:1992ft,Becirevic:2003wk,AliKhan:2003cu,Arndt:2004bg,Beane:2004tw,Beane:2004rf,Bedaque:2006yi},
but as far as we know the only finite volume two--loop calculation which
has been performed until now is for the quark
condensate~\cite{Bijnens:2005ne}, which is a much simpler calculation. In
the related context of finite temperature field theory there are a few
examples of calculations beyond one loop~\cite{Schenk:1993ru,Toublan:1997rr}.

We wish to briefly mention related work. Most notably, the asymptotic
formula may also be applied to the nucleon mass \cite{Luscher:1983rk}, see
Koma and Koma~\cite{Koma} as well as \cite{Colangelo:2005cg} for recent
work. Braun, Pirner and Klein have evaluated the volume dependence of the
pion mass based on a quark--meson model \cite{Braun:2004yk}. In the
framework of a lattice regularised ChPT finite volume effects have been
addressed by Borasoy et al. in Ref.~\cite{Borasoy:2005nz}.

The outline of the article is as follows. In sect.~\ref{sec:preli} we set
the notation and remind of the basic assumptions for an application of ChPT
in finite volume. Sect.~\ref{sec:result} is devoted to outline the
calculation and state the main results. In sect.~\ref{sec:analytical} we
show the explicit expressions which have been used for the numerical
analysis in sect.~\ref{sec:numerical}. We conclude with a summary.

\section{Preliminaries}
\label{sec:preli}

\noindent
In this section we shall set the notation and the basic definitions for the
two--loop calculation. 

\subsection{ChPT in finite and in infinite volume}
\label{sec:chpt}

\noindent
Chiral Perturbation Theory (ChPT) is the effective theory for QCD at low
energies. If we first restrict ourselves to the infinite volume case, the
effective Lagrangian of QCD for two light flavours at low energies consists
of an infinite number of terms \cite{Gasser:1983yg},
\begin{equation}
  \lagr_\mr{eff} = \lagr_2 + \lagr_4 + \lagr_6 + \ldots \, .
\end{equation}
As we wish to calculate the pion mass, an on--shell quantity,
external fields can be dropped in $\lagr_\mr{eff}$. We work
in the isospin symmetry limit $m_u=m_d$ in Euclidean space--time, and for the
choice of the pion fields we use the non--linear sigma model parameterization.
We have 
\begin{equation}
  \lagr_2 = \frac{F^2}{4} \left\< u_\mu u_\mu-\chi_+\right\>\co
\end{equation}         
with
\begin{eqnarray}         
  U &=& \sigma + i \frac{\mbox{\boldmath$\phi$}}{F} \qquad , \qquad
\sigma^2 + \frac{\mbox{\boldmath$\phi$}^2}{F^2} = \unith \co \; \;
\mbox{\boldmath$ \phi$} =
 \left( \mbox{$\begin{array}{cc} \pi^0
& \sqrt{2} \pi^+
         \\ \sqrt{2} \pi^- &- \pi^0 \end{array}$} \right) =
\phi^i \tau^i  \co\nn
         u_\mu &=& i u^\dagger \partial_\mu Uu^\dagger = -i u
\partial_\mu U^\dagger u = u_\mu^\dagger \co \; \; 
         \chi_+ = u^\dagger \chi u^\dagger + u \chi^\dagger u
\co \nn
         \chi &=& 2 B \hat{m}\unith \qquad , \qquad
\hat{m}=\frac{1}{2}(m_u+m_d)\co
\label{eq:sigma_param}
\end{eqnarray}
with $u^2=U$. The symbol $\< A \>$
denotes the trace of the two--by--two matrix $A$.
The Lagrangian $\lagr_4$ can be written as \cite{Gasser:1983yg},
\begin{equation}
         \lagr_4 = \sum^{4}_{i=1} \ell_iP_i + \dots \co
         \label{eqleff4}
\end{equation}
         where
\begin{equation}
         P_1 = -\frac{1}{4} \< u_\mu u_\mu \>^2 \, , \;
         P_2 = -\frac{1}{4} \< u_\mu u_\nu\>^2 \, , \; 
         P_3 = -\frac{1}{16} \< \chi_+\>^2 \, , \;
         P_4 =  \frac{i}{4} \< u_\mu \chi_{-\, \mu} \> \co
\end{equation}
with
\begin{equation}
  \chi_{-\, \mu} = u^\dagger \partial_\mu \chi u^\dagger
                 -u\partial_\mu \chi^\dagger u \fs
\end{equation}
The ellipsis in eq.~(\ref{eqleff4}) denotes terms that do not
contribute to the pion mass.  The low energy
constants $\ell_i$ are divergent and remove the ultraviolet
divergences generated by one--loop graphs from ${\cal L}_2$. 

The complete effective Lagrangian $\lagr_6$ with its divergence structure
at $d=4$ has been constructed in \cite{Bijnens:1999sh,Bijnens:1999hw}. As
will be discussed in sect.~\ref{sec:result}, terms from $\lagr_6$ merely
renormalize the pion mass in infinite volume and do not contribute to
finite volume corrections which we are interested in. Thus, we refrain from
showing it here. Given the effective Lagrangian and the parameterization
for the pion fields, it is straightforward to calculate the pion mass to
two--loops. We refer to \cite{Bijnens:1997vq}, where one also finds a
detailed discussion of the renormalization procedure [two--loop diagrams in
ChPT are discussed in~\cite{Gasser:1998qt}].

The effective framework is still appropriate, when the system is enclosed
by a large box of size $V=L^3$. We refer to the literature for the
foundations~\cite{Gasser:1986vb,Gasser:1987ah,Gasser:1987zq} and a recent
review~\cite{Colangelo:2004sc}. Here, we only recall a few fundamental
results which guided the present calculation: the volume has to be large
enough, such that ChPT can give reliable results, $2\fpi L \gg 1$. The
value of the parameter $\mpi L$ determines the power counting for the
perturbative calculation: if $\mpi L \gg 1$ one is in the ``$p$--regime''
in which $1/L$ counts as a small quantity of order $M_\pi$. If $\mpi L
\lesssim 1$ one is in the ``$\epsilon$--regime'' and $1/L^2$ is a quantity
of order $M_\pi$. In both cases the effective Lagrangian is the same
as in the infinite volume. In this article we only consider the
``$p$--regime'', where the system is distorted mildly and the only change
brought about by the finite volume is a modification of the pion propagator
due to the periodic boundary conditions of the pion
fields\footnote{Throughout we denote by the volume the three--dimensional
  volume $V=L^3$, whereas the time direction is not compactified.}
\begin{equation}
G(x^0,\mb{x})=\sum_{{\mb n} \in \mathbb{Z}^3}G_0(x^0,{\mb x}+{\mb n}L) \, ,
\label{eq:Gx1}
\end{equation}
with $G_0(x)$ the propagator in infinite volume.

\subsection{Basic definitions}

\noindent
In Euclidean space--time the propagator is defined through the connected
correlation function
\begin{equation}
  G(x)\delta^{ab} = \<\phi^a(x)\phi^b(0)\>_L \, ,
\end{equation}
where $a$,$b$ stand for isospin indices of the pion fields and the subscript
$L$ in the correlation function denotes that it is evaluated in finite
volume. 

\noindent
We also have 
\begin{eqnarray}
  \<\phi^1(x)\phi^1(0)\>_L &=& L^{-3}\sum_{\bf p} \int \frac{dp^0}{2\pi} 
                             e^{ipx}\, G(p^0,{\bf p}) \, ,\nn
             G(p^0,{\bf p})^{-1} &=& M^2+p^2-\Sigma_L(p^0,{\bf p}) \, ,
\end{eqnarray}
where the momenta ${\bf p}$ can only have discrete values
\begin{equation}
  {\bf p} = \frac{2\pi}{L}{\bf n} \, , \qquad {\bf n} \in \mathbb{Z}^3 \, ,
\end{equation}
and $M^2=2B\hat{m}$ is the tree--level pion mass in infinite volume.
The pion mass in finite volume $\mpiL$ is now defined by the pole equation
\begin{equation}
  \label{eq:pole}
  G(\hat{p}_L)^{-1} = 0 \co \qquad \mbox{for} \qquad 
  \hat{p}_L = (i\mpiL,{\bf 0}) \fs
\end{equation}

\section{Outline of calculation and statement of results}
\label{sec:result}

\noindent
For a large volume, the finite size effects are expected to
be small, such that the pole equation can be solved perturbatively. We outline
the calculation and proceed with the main results. More details about the
calculation are relegated to later sections or the appendix.

\subsection{One--loop result}

\begin{figure}[t]
\begin{center}
\includegraphics[width=10cm]{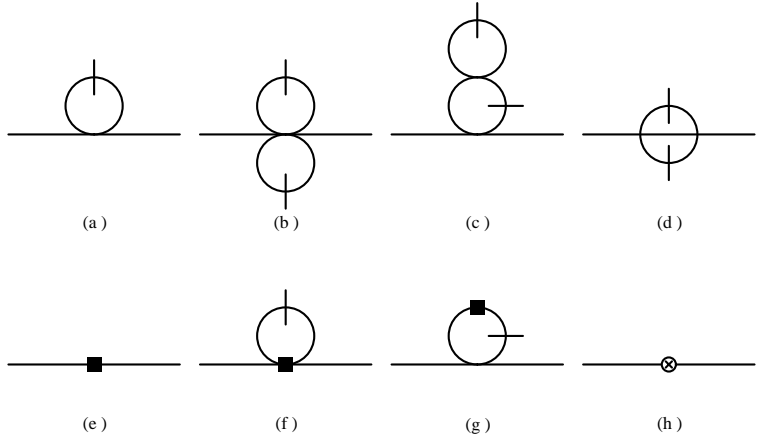} 
\end{center}
\caption{Self--energy graphs to two--loops in ChPT. A spline corresponds
  to a periodified propagator, whereas those without correspond to an
  infinite volume propagator, cf. eq.(\ref{eq:Gx1}). Normal vertices come
from $\lagr_2$, squared vertices from $\lagr_4$ and the circle--crossed from
$\lagr_6$.
\label{fig:selfe}}
\end{figure}

\noindent
Since the effective Lagrangian remains unchanged when going to the finite
volume, we can immediately write down the Feynman diagrams which contribute
to the self--energy at two--loops, see fig.~\ref{fig:selfe}, the only
difference with respect to an infinite volume calculation is that the
propagators need to be periodified, cf. eq.(\ref{eq:Gx1}).
At leading order, the graphs $1(a)$ and $1(e)$ need to be evaluated and the
self--energy admits the form
\begin{equation} 
  \label{eq:se_oneloop}
  \Sigma_L(p^0,{\bf p}) = \frac{1}{\fpi^2}G(0)
  \left(-\frac{3}{2}\mpi^2-p^2\right)
  -2\ell_3\frac{\mpi^4}{\fpi^2}
  +\cO\left(\frac{1}{\fpi^4}\right) \co
\end{equation} 
with $G(0)$ the value of the finite volume propagator at the origin. It
contains a logarithmic divergence due to the contribution from the
term ${\bf n=0}$ in eq.(\ref{eq:Gx1}). In dimensional regularization, 
\begin{equation} 
  \label{eq:G0} 
  G_0(0) = \frac{1}{(2\pi)^d}\int d^dp
  \frac{1}{p^2+M^2} 
  = \frac{\Gamma(1-d/2)}{(4\pi)^{d/2}}M^{d-2} \fs
\end{equation} 
The remaining terms with ${\bf n\neq0}$ are finite and may be expressed in
terms of a kinematical function\footnote{The second argument of $g_1$ is
  the temperature, which we keep zero. Notation as in
  Ref.\cite{Gasser:1986vb}} $g_1(M^2,0,L)$,
\begin{eqnarray} 
  G(0) &=& G_0(0) + g_1(M^2,0,L) \, ,\nn
  g_1(M^2,0,L) &=& \int_0^\infty \frac{d\tau}{(4\pi\tau)^{d/2}}
  e^{-\tau M^2}
  \sum_{\bf n\neq 0}
  \exp\left(-\frac{{\bf n^2}L^2}{4\tau}\right)
  \label{eq:g1dec}
\end{eqnarray}
For a derivation of eq.(\ref{eq:g1dec}), we refer to
\cite{Hasenfratz:1989pk}. In app.~\ref{app:1} we provide a different
derivation, based on a contour integration analysis. The pole of the Gamma
function in eq.(\ref{eq:G0}) in four dimensions is absorbed in a
renormalization of the low--energy constant $\ell_3$. One readily verifies
that inserting eq.(\ref{eq:g1dec}) into eq.(\ref{eq:se_oneloop}) yields for
the leading finite volume shift \cite{Gasser:1986vb}
\begin{equation}
\mpiL = \mpi\left[ 1+\frac{1}{4\fpi^2}g_1(\mpi^2,0,L)
  +\cO\left(\frac{1}{\fpi^4}\right) \right] \fs
\end{equation}
The separation of the cut--off and the volume--dependence is as expected:
finite volume corrections do not generate new uv--divergences. At leading
order the finite volume corrections could be isolated immediately. This
will not be the case at the two--loop level. The graph $(d)$ in
fig.~\ref{fig:selfe} does not factorize in pure one--loop integrals and
further steps need to be performed. We refer to app.~\ref{app:1} for
further details.

\subsection{Minimal set of periodified propagators}
\label{sec:minimal}

\noindent
In the evaluation of the Feynman diagrams, it is mandatory to make use of
the following result. In order to evaluate a graph consisting of $L$ loops,
one needs to consider only a certain set of $L$ propagators as the
periodified, finite-volume ones -- the others can be taken as
infinite-volume propagators. In the following we briefly discuss
this statement (see also p.18ff in Ref.\cite{Luscher:1985dn}).  Consider an
arbitrary self--energy graph with $L$ loops, $I$ internal lines $\ell$ and
$V$ vertices. Since the number of loops is the number of independent
integrations over momenta, we have
\begin{equation}
  L = I-V+1 \, .
\end{equation}
For every line $\ell$ the propagator is an infinite sum of terms
characterized by an integer vector $\mb{n}(\ell)$. The loop graph itself is
an infinite sum of terms which can be identified by a set of integer
vectors $\{\mb{n}(\ell)\}$ assigned to all the internal lines. As remarked
by L\"uscher this structure can be viewed as a gauge field on the
self--energy graph. Gauge transformations are defined as\footnote{We borrow
  the notation from Ref.\cite{Luscher:1985dn}.}
\begin{equation}
  \mb{n}'(\ell) = \mb{n}(\ell) 
  + \mb{\Lambda}(f(\ell))-\mb{\Lambda}(i(\ell)) \, ,
\end{equation}
where $\mb{\Lambda}(v)$ is some field of integer vectors and $i(\ell)$
($f(\ell)$) is the initial (final) vertex of the line $\ell$. One verifies
immediately that two contributions of a Feynman graph which differ only by
a gauge transformation yield the same mathematical expression. In the
calculation of the loop graph what matters is the sum over representatives
of gauge equivalence classes. A convenient representative can be found,
e.g., by adjusting $\mb{\Lambda}(i(\ell))$ and $\mb{\Lambda}(f(\ell))$
iteratively, such that $\mb{n}(\ell)=0$ for as many internal lines $\ell$
as possible.  This can be achieved for $V-1$ lines. Therefore, there remain
$I-V+1$ internal lines where the periodified propagator has to be inserted
-- which coincides with the number of loops of the graph. This shall be our
minimal set of periodified propagators. In fig.~\ref{fig:selfe}, we have
attached a spline to a periodified propagator, whereas lines without a
spline correspond to an infinite volume propagator.

\subsection{Two--loop result}
\label{sect:two--loop}

\noindent
It is convenient to split the sum over the equivalence classes into three
parts\footnote{Note the slight difference in the definition of simple
  fields with respect to \cite{Luscher:1985dn}. We do not require
  $|\mb{n}(\ell)|=1$.}:

\vspace*{0.5cm}
\parbox{\textwidthMinusEight}{
\begin{tabular}{ll}
  $\Sigma^{(0)}$: & $\mb{n}(\ell)=0$ $\forall$ $\ell$ (pure gauge),\\
  $\Sigma^{(1)}$: & $\mb{n}(\ell)=0$ $\forall$ $\ell$ except for one line 
  $\bar\ell$ (simple gauge),\\
  $\Sigma^{(2)}$: & $\mb{n}(\ell)=0$ $\forall$ $\ell$ except for two lines
  $\bar\ell_1$,$\bar\ell_2$. 
\end{tabular}}
\parbox{8mm}{
  \begin{equation}
    \label{3tab:split}
  \end{equation}}
\vskip 2mm

\noindent
The pion mass in finite volume to two--loops then admits the form
\begin{eqnarray}
  \label{eq:sum1}
  \mpiL^2 &=& \mpi^2 -\Sigma^{(1)}- \Sigma^{(2)} \co \nn
  \mpi^2  &=& M^2 - \Sigma^{(0)} \co
\end{eqnarray}
where, using  $\lapi = \mpi L$, we get
\begin{eqnarray}
  \label{eq:sum2}
  \Sigma^{(1)} &\!\!=\!\!& I_p + I_c + \cO(\xi^3) \co \\[1.5ex]
  \label{eq:res_luscher}
    I_p        &\!\!=\!\!&
    {\mpi^2\over16\pi^2\lapi}\;\sum_{n=1}^{\infty}{m(n)\over\sqrt{n}} 
  \int\limits_{-\infty}^\infty\!dy\;
  \cF_\pi(\ri y)\,e^{-\sqrt{n(1+y^2)}\lapi} \co \\[1.5ex]
    I_c        &\!\!=\!\!&
  -\frac{\ri\mpi^2}{32\pi^3\lapi} 
  \sum_{n=1}^{\infty} \frac{m(n)}{\sqrt{n}} \!
  \int\limits_{-\infty}^{\infty} \! dy
  \int\limits_{4}^{\infty} d\til s\,
  \frac{e^{-\sqrt{n(\til s+y^2)}\lapi}}
  {\til s+2\ri y}\,
  \mr{disc}\big[\cF_\pi(\til s,1+\ri y)\big] \fs 
  \label{eq:Ic1}
\end{eqnarray}
with $m(n)\equiv$ number of integer vectors $\bf{z}$ with ${\bf{z}}^2=n$.
$I_p$ denotes the pole and $I_c$ the cut contribution, whose meaning and
precise definitions will be explained below. The expression for
$\Sigma^{(2)}$ is more cumbersome:
\begin{equation}
  \label{eq:Sig2}
  \Sigma^{(2)} = \mpi^2\,
  \xi^2 \bigg[ 
                \frac{9}{8}\,       \til{g}_1(\lambda_\pi)^2
               -\frac{1}{8}\lapi\, \til{g}_1(\lambda_\pi)
                \frac{\partial}{\partial \lapi} \til{g}_1(\lapi)
               +\Delta
               \bigg] 
               +\cO(\xi^3) \co 
\end{equation}
and we merely note that it can be split into products of one-loop
contributions (terms with $\til{g}_1^2$ and a pure two-loop part, denoted
by $\Delta$. We have introduced the abbreviation
\begin{equation}
  \label{eq:xi}
  \xi = \frac{\mpi^2}{16\pi^2 \fpi^2} \fs
\end{equation}
A detailed derivation of these results will be given in the subsequent
sections and we confine ourselves to a few comments at this stage: In $I_p$
one recovers the asymptotic formula of L\"uscher, if one restricts the sum
to the first addendum. Its extension to the present form has already been
suggested in~\cite{Colangelo:2004sc} and was applied
in~\cite{Colangelo:2005gd}. The function $\cF_\pi(\nu)$ denotes the isospin
zero $\pi\pi$ scattering amplitude in the forward kinematics. It contains
cuts due to the two--pion intermediate state. In the derivation of the
asymptotic formula, L\"uscher consistently dropped contributions arising
from the cuts, since in a large volume expansion they are beyond the order
of accuracy he aimed at. Here, we take them into account up to two--loops,
relying on the chiral representation of $\pi \pi$ scattering amplitude. The
outcome of the analysis is summarized in $I_c$ in which the same symbol
$\cF_\pi$, this time with two arguments appears: with the latter we mean
the scattering amplitude of two on-shell pions into two off-shell pions in
the forward kinematics configuration. The contributions from the cuts can
still be written as integrals over the $\pi\pi$ scattering amplitude with
an exponential weight, as in the asymptotic formula. Notice however, that
this formula explicitly relies on the chiral representation of the
scattering amplitude and is only valid up to two--loops, whereas the
L\"uscher formula holds to all orders.

Contributions from two pion propagators in finite volume are ultimately
captured in $\Sigma^{(2)}$, being expressed in terms of a dimensionless
function $\til{g}_1(\lapi)$ and a numerically small correction $\Delta$
arising from graph \ref{fig:selfe}\,d). Both are explicitly given in
app.~\ref{app:1} and eq.~(\ref{eq:Delta}), respectively.

The pure gauge contributions are not volume dependent and merely renormalize
the pion mass, cf. eq.~(\ref{eq:sum1}). A detailed discussion of this
calculation can be found in \cite{Burgi:1996qi,Bijnens:1997vq}, with
which we agree -- this was a useful, nontrivial check for our calculation.

\subsection{Self--energy to first order: $\Sigma^{(1)}$}
\label{sec:sig1}
\begin{figure}[t]
\begin{center}
\includegraphics[width=8cm]{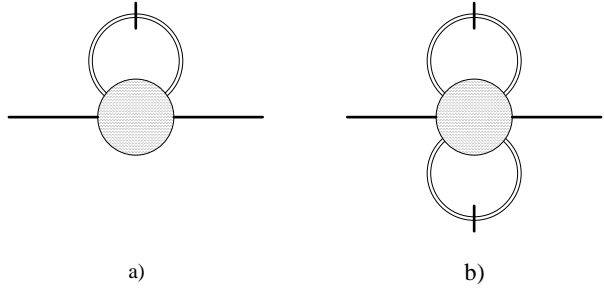}
\end{center}
\caption{Skeleton diagrams representing $a)$ eq.~(\ref{eq:sig1}) and
  $b)$ eq.~(\ref{eq:self2}). The blob in $a)$ stands for the 4--point function
  of $\pi\pi$ scattering in infinite volume and in $b)$ for a subtracted
  6--point function of $\pi\pi\pi$ scattering in infinite volume. The
  double--line with the spline is a finite volume propagator with the physical
  pion mass $\mpi^2$.\label{fig:skel}} 
\end{figure}

\noindent
The simple fields can be summed up in closed form and may be represented
by a skeleton diagram, see fig.~\ref{fig:skel}$a)$
\begin{equation}
  \label{eq:sig1}
  \Sigma^{(1)} = \frac{1}{2}\int\frac{d^4q}{(2\pi)^4}
  \sum_{n=1}^{\infty} m(n) \, 
  e^{iq_1\sqrt{n}L} G_0(q^2)
  \,\Gamma_{\pi\pi}(\hat{p},q,-\hat{p},-q) \co
\end{equation}
with $\hat{p} = (i\mpi,{\bf 0})$ and
$\Gamma_{\pi\pi}(\hat{p},q,-\hat{p},-q)$ the 4--point function of $\pi\pi$
scattering in the forward scattering kinematics. Note that the 4--point
function in eq.~(\ref{eq:sig1}) is an off--shell amplitude and that we
evaluate it at $\hat{p}$ and not at $\hat{p}_L$. The difference is taken
into account in $\Sigma^{(2)}$. The representation of $\Sigma^{(1)}$ in
eq.~(\ref{eq:sig1}) has already been used by L\"uscher and eventually led
him to the asymptotic formula \cite{Luscher:1985dn}.
He then discussed the contribution of the pole of the propagator
$G_0(q^2)$ that one meets at  
\begin{equation}
  \label{eq:pole_q1}
  q_1 = i\sqrt{\mpi^2+q_\bot^2 + q_0^2} \, , \qquad 
  q_\bot = (q_2,q_3) \, .
\end{equation}
Above this pole the singularities come from the cuts of the propagator and the
4--point function $\Gamma_{\pi\pi}(\hat{p},q,-\hat{p},-q)$. These start from  
\begin{equation}
       s =  -(\hat{p}+q)^2 \geq 4\mpi^2 \co
\quad u =  -(\hat{p}-q)^2 \geq 4\mpi^2 \co
\quad            -q^2 \geq 9\mpi^2 \, .
\end{equation}
L\"uscher showed that the pole contribution is dominating with respect to
those coming from the cuts and neglected the latter.
In fact, his discussion involved no further assumptions about the
4--point function and remains true at every order of the perturbative
expansion. Since our goal is to test the asymptotic formula beyond the leading
exponentials, we wish to work out the impact of the contributions that
were dropped by L\"uscher. Doing this at the two--loop level is rather
straightforward, as we will now show.

\begin{figure}[t]
\begin{center}
\includegraphics[width=6cm]{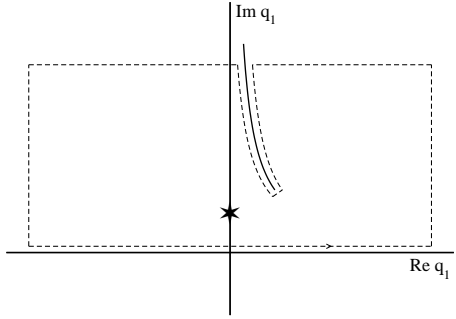}
\end{center}
\caption{Integration contour in the complex $q_1$ plane with the pole
  from the pion propagator and the branch cut from the $\pi\pi$ scattering
  amplitude.} 
\label{3fig:cut}
\end{figure}

Up to $\cO(p^4)$, the four--point function can be decomposed into a
combination of functions which have either a singularity in $s$ or in $u$.
Since $\Sigma^{(1)}$ is symmetric in $s$ and $u$, we may write
$\Gamma_{\pi\pi}(\hat{p},q,-\hat{p},-q)$ in terms of a function
$\bar{\Gamma}_{\pi\pi}(s,\hat{p}q)$, which depends only on the variables
$s$ and $\hat{p}q$, and which has cuts at $s\geq 4\mpi^2$,
\begin{equation}
  \Gamma_{\pi\pi}(\hat{p},q,-\hat{p},-q) = 
  \bar{\Gamma}_{\pi\pi}(s,\hat{p}q) + \cO\left(\frac{1}{\fpi^6}\right)
  \, .
\end{equation}
We write $q_1 = x+i y$ and for the domain $s \in \mathbb{R}$, $s\geq
4\mpi^2$ we have 
\begin{equation}
  y^2-x^2-q_0^2-q_\bot^2 \geq 3\mpi^2 \, , \qquad xy = -q_0 \mpi \, .
\end{equation}
In particular we observe that in the complex $q_1$ plane the cut starts
above the pole of eq.~(\ref{eq:pole_q1}), as illustrated in
fig.~\ref{3fig:cut}. 
Performing now the contour integration in the upper half plane we obtain
two terms: the first one, $I_p$, comes from the pole and is simply its
residuum. The other comes from the integral along the new integration path.
The latter contribution vanishes as we push the integration lines to
infinity, except for the path which goes around the cut, to be denoted by
$I_c$ in the following,
\[
  \Sigma^{(1)}   =  I_p + I_c 
  \, .
\]
As was shown by L\"uscher~\cite{Luscher:1985dn}, the former can be
simplified considerably and can be brought in the usual form of the
asymptotic formula
\[
    I_p        =
    {\mpi^2\over16\pi^2\lapi}\;\sum_{n=1}^{\infty}{m(n)\over\sqrt{n}} 
  \int\limits_{-\infty}^\infty\!dy\;
  \cF_\pi(\ri y)\,e^{-\sqrt{n(1+y^2)}\lapi} \co
\]
with $\cF_\pi(iy)$ the $\pi\pi$ forward scattering amplitude.  Restricting
the sum to the first term we recover L\"uscher's formula
\cite{Luscher:1985dn}.  For $I_c$ we find
\begin{equation}
  I_c = \frac{1}{2}
  \sum_{n=1}^{\infty} m(n) \,
  \int \frac{dq_0 d^2q_\bot}{(2\pi)^4} 
  \hspace*{-0.2cm} \int\limits_{[s\geq 4\mpi^2]}\hspace*{-0.4cm} dq_1
  \frac{e^{iq_1\sqrt{n}L}}{\mpi^2+q^2}\;
  \mr{disc}\big[\bar\Gamma_{\pi\pi}(s,\hat{p}q)\big] 
  +\cO(\fpi^{-6})
  \, , 
\end{equation}
where $\mr{disc}\big[\bar\Gamma_{\pi\pi}(s,\hat{p}q)\big]$ denotes the
discontinuity of $\bar\Gamma_{\pi\pi}$ along the cut. 
It is now convenient to shift the integration path in $q_0$ from
$\mathrm{Im}(q_0)=0$ to $\mathrm{Im}(q_0) =-\ri\mpi$. Along this path we
have 
\begin{equation}
\label{eq:off--shell}
  q_0 = \bar{q}_0 -\ri\mpi \, ,\qquad 
    s = -\bar{q}_0^2 - q_1^2 - q_\bot^2 \, ,\qquad 
  \bar{q}_0 \in \mathbb{R} \, ,
\end{equation}
and the integration over $q_1=x+\ri y$ falls onto the imaginary axis,
\begin{eqnarray}
  x  &=& 0 \, ,\qquad y \geq y_0 = \sqrt{4\mpi^2+\bar{q}_0^2+q_\bot^2} \co\nn 
 I_c &=& \frac{\ri}{2}
 \sum_{n=1}^{\infty} m(n) \,
 \int \frac{d\bar{q}_0 d^2q_\bot}{(2\pi)^4} 
 \int\limits_{y_0}^\infty
  dy\,  
  e^{-y\sqrt{n}L}\, \frac{\mr{disc}
  \big[\bar\Gamma_{\pi\pi}(s,\hat{p}q)\big]}{\mpi^2+q^2}
  \fs
\end{eqnarray}
Next, we change the integration variable from $q_1$ to $s$,
\begin{equation}
 I_c = -\frac{\ri}{2}
 \sum_{n=1}^{\infty} m(n)\,
 \int \frac{d\bar{q}_0 d^2q_\bot}{(2\pi)^4} \!\!
 \int\limits_{4\mpi^2}^\infty \!ds\,
 \frac{e^{-\sqrt{n(s+\bar{q}_0^2+q_\bot^2)}L}}{2(s+\bar{q}_0^2+q_\bot^2)^{1/2}}
    \, \frac{\mr{disc}
 \big[\bar\Gamma_{\pi\pi}(s,\hat{p}q)\big]}{s+2\ri \mpi\bar{q}_0} 
  \, ,
\end{equation}
and make use of 
\begin{equation}
  \int \frac{d^2q_\bot}{(2\pi)^2}\frac{1}{2(\mu^2+q_\bot^2)^{1/2}}
  e^{-\sqrt{n(\mu^2+q_\bot^2)}L} = 
  \frac{1}{4\pi\sqrt{n}L}e^{-\mu\sqrt{n}L}\, , 
\end{equation}
to carry through the integration over $q_\bot$ and to end up with
eq.~(\ref{eq:Ic1}).

\subsection{Self--energy to second order: $\Sigma^{(2)}$}

The term $\Sigma^{(2)}$ is more complicated to manipulate, although it is
suggestive to think that it may be related to an integral over the $3 \pi
\to 3 \pi$ amplitude. A similar question has been asked in the framework of
finite temperature QCD~\cite{Schenk:1993ru}.  We discuss the situation in
finite volume in appendix~\ref{app:sigma2}. Since these considerations have
not lead us to a nice and compact representation for $\Sigma^{(2)}$ we have
written it as in (\ref{eq:Sig2}) where we only have split the factorizable
two--loop contributions from the rest. The latter, which we have denoted by
$\Delta$ we have evaluated numerically.

We conclude this section with a remark concerning the large $L$ behaviour
of the two--loop results. In the large volume limit, the contributions from
eq.~(\ref{eq:res_luscher}) behave according to
\begin{equation}
  \label{eq:largeL}
  \lim_{L\rightarrow\infty}I_p \sim 
  \frac{1}{\lapi^{3/2}}\,e^{-\lapi} \fs
\end{equation}
Similarly, we may evaluate the large volume behaviour of the terms
occurring in eqs.(\ref{eq:Sig2},\ref{eq:Ic1}). It turns out that these,
besides being exponentially suppressed and behaving at least as $\exp(-
\alpha \lapi)$, with $\alpha=(\sqrt{3}+1)/\sqrt{2}$, are also suppressed by
a power of $\lapi$. Unfortunately this suppression is not particularly
strong, being of $1/\sqrt{\lapi}$ with respect to eq.~(\ref{eq:largeL}).
While we can conclude that the resummed L\"uscher formula is dominating in
comparison to other two--loop diagrams in a $1/\sqrt{\lapi}$ expansion,
i.e.
\begin{eqnarray}
  \label{eq:poly_mpiL}
    \mpiL     &=& \bar\mpiL (1+\cO ( e^{-\alpha\lapi}/ \sqrt{\lapi})) \co \nn
    \bar\mpiL &=& \mpi - \frac{1}{2\mpi}I_p \co
\end{eqnarray}
we are convinced that the most important test of the resummed formula is
the numerical one, which we will discuss in the following.

\section{Summary of analytical results}
\label{sec:analytical}

\noindent
In this section we shall give the analytical results in explicit form, i.e.
insert the chiral representation of the amplitudes which appear in the
formulae given in the previous section and express our results in terms
of a few basic integrals. In eq.(\ref{eq:sum1}--\ref{eq:xi}) we have split
the finite size effects of the pion mass into $\Sigma^{(1)}= I_p+I_c$ and
$\Sigma^{(2)}$. For the former two we find
\begin{eqnarray}
  I_p &=& \mpi^2 \sum\limits_{n=1}^{\infty} \frac{m(n)}{\sqrt{n}}
          \frac{1}{\lapi}\xi \left[ I_{\mpi}^{(2)} + \xi I_{\mpi}^{(4)}
           \right] \co \nn
  I_c &=& \mpi^2 \sum\limits_{n=1}^{\infty} \frac{m(n)}{\sqrt{n}}
          \frac{1}{\lapi}\xi^2 \til I_c \co 
\label{eq:Ic}
\end{eqnarray}
where $\xi$ is the chiral expansion parameter defined in eq.(\ref{eq:xi})
and the expressions $I_{\mpi}^{(2)}$, $I_{\mpi}^{(4)}$ have already been
given in Ref.~\cite{Colangelo:2003hf,Colangelo:2005gd}, and we reproduce
them here for completeness:
\begin{eqnarray}
I^{(2)}_{M_\pi}&=&-B^0
\\
I^{(4)}_{M_\pi}&=&B^0
\bigg[ -\frac{55}{18}+4\lb_1+\frac{8}{3}\lb_2-\frac{5}{2}\lb_3-2\lb_4 \bigg]
+B^2 \bigg[ \frac{112}{9}-\frac{8}{3}\lb_1-\frac{32}{3}\lb_2 \bigg] \nn
&+&\frac{13}{3}R_0^0-\frac{16}{3}R_0^1-\frac{40}{3}R_0^2 \co \nonumber
\end{eqnarray}
where
\begin{equation}
B^{0,2}\equiv B^{0,2}(\sqrt{n}\la_\pi)\;, \qquad
B^0(x)=2K_1(x)\;,\qquad
B^2(x)=2K_2(x)/x \co
\end{equation}
and the integrals $R_0^k$ are defined as 
\begin{eqnarray}
R_0^{k}&=&
\left\{{\mr{Re}\atop\mr{Im}}\right.
\int_{-\infty}^{\infty}\!d\til y\;\til y^k\,e^{-\sqrt{n(1+\til y^2)}\,\la_\pi}\,
g(2+2\ri\til y)\qquad\;\;\;\;
\mr{for}\;\left\{{k\;\mr{even}\atop k\;\mr{odd}}\right. \co
\label{eq:R0k}
\end{eqnarray}
where $g(x)$ is related to the standard one--loop function $\bar{J}(x\mpi^2)$
through
\begin{equation}
  g(x) = 16\pi^2 \bar{J}(x\mpi^2) \co
\end{equation}
and $\bar{J}(x\mpi^2)$ given in eq.~(\ref{eq:H1}). For
instance for $x<0$, we have 
\begin{equation}
g(x)=\sigma \log \frac{\sigma -1}{\sigma+1} + 2 \;, \qquad \mbox{with}
\qquad \sigma=\sqrt{1-4/x} \co
\end{equation}
and elsewhere defined through analytic continuation. In
Ref.\cite{Colangelo:2005gd} also the next-to-next-to-leading order term in
$I_p$, namely $ \xi^2 I_{\mpi}^{(6)}$ has been given and we will use it in
our numerical analysis.  The coefficient $\til I_c$ can be expressed as a
combination of basic integrals
\begin{eqnarray}
  \til I_c &=&  \frac{1}{3}
  \left(
  112 C^{0,2} + 37 C^{1,0} - 40 C^{1,2} - 4 C^{2,0}
  \right) \co \nn
  C^{j,k} &=& 
  \int\limits_{-\infty}^{\infty} dy
  \int\limits_4^\infty d\til s\,\,
  \frac{e^{-\sqrt{n(\til s+y^2)}\lapi}}{\til s^2+4y^2}
  \left(1-\frac{4}{\til s} \right)^{1/2}\,
  \til s^j y^k \fs
\label{eq:til_Ic}
\end{eqnarray}
Notice that the latter expression is obtained from the chiral
representation of the off-shell $\pi \pi$ scattering amplitude for forward
kinematics. This is parameterization dependent, and the result given here is
obtained for the parameterization discussed in sect.~\ref{sec:chpt}. The
dependence on the parameterization must cancel in the full result.

\noindent
The self--energy to second order has already been introduced
in eq.(\ref{eq:Sig2}) and reproduced here for convenience
\[
  \Sigma^{(2)} = \mpi^2\,
  \xi^2 \bigg[ 
                \frac{9}{8}\, \til{g}_1(\lambda_\pi)^2
               -\frac{1}{8}\lapi\, \til{g}_1(\lambda_\pi)
                \frac{\partial}{\partial \lapi} \til{g}_1(\lapi)
               +\Delta
               \bigg] 
               +\cO(\xi^3) \co
\]
with
\begin{equation}
\label{eq:Delta}
  \Delta = (16\pi^2)^2 \Big[4\til p_\mu \til p_\nu H_{\mu\nu} 
  + 4\til p_\mu H_\mu + \frac{7}{6}H \Big] \co 
  \qquad \til p = \frac{\hat p}{\mpi} \co
\end{equation}
where $H$,$H_{\mu}$ and $H_{\mu\nu}$ are related to the 
sunset--type integrals of fig.~\ref{fig:selfe}$d)$. We have not been able to
find a compact representation for these integrals and only elaborate on their
numerical analysis in app.~\ref{app:1} in some detail. 

\section{Numerics}
\label{sec:numerical}

\subsection{Setup}
\noindent
The numerical analysis is performed in line with the setup of
Ref.\cite{Colangelo:2005gd}. The quantity of interest is
\begin{equation}
  R_{\mpi} \equiv \frac{\mpiL-\mpi}{\mpi} \co
\end{equation}
whose quark mass dependence shall be evaluated numerically for different
sizes $L$. The parameters of $R_{\mpi}$ are (see
eqns.(\ref{eq:sum1}--\ref{eq:xi}) and eqns.(\ref{eq:Ic}--\ref{eq:Delta}))
the pion mass $\mpi$ and the pion decay constant $\fpi$ in infinite volume
as well as (implicitly in $I_{\mpi}^{(4/6)}$) the SU(2) low energy
constants. The quark mass dependence of the pion decay constant may be
taken into account by expressing $\fpi$ as a function of the pion mass
$\mpi$ (see, e.g.~\cite{Colangelo:2003hf}).  Regarding the low energy
constants, we use the ones determined in
\cite{Bijnens:1997vq,Colangelo:2001df} which are the same as in our
previous finite size studies \cite{Colangelo:2003hf,Colangelo:2005gd}.

\subsection{Results}

\begin{figure}[t]
\begin{center}
\includegraphics[width=12cm]{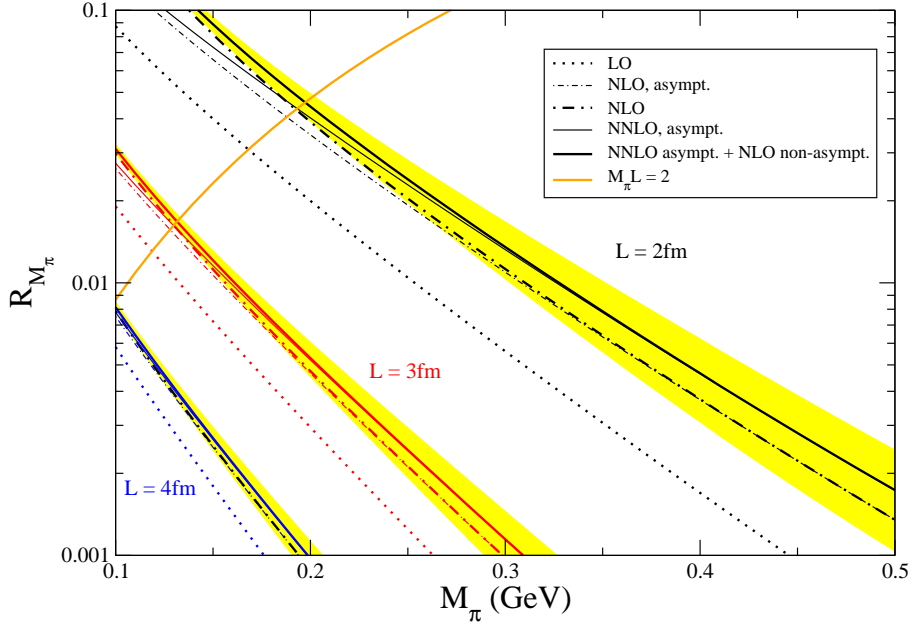}
\end{center}
\caption{$R_{\mpi}=\mpiL/\mpi-1$ vs. $\mpi$ for $L=2,3,4$fm. 
  The result of the resummed
  asymptotic L\"uscher formula (\ref{eq:res_luscher}) with LO/NLO/NNLO chiral
  input (with attribute ``asympt.'' for NLO and NNLO in legenda) is compared to
  the one--loop (LO) and two--loop (NLO) result. The best estimate for
  $R_{\mpi}$ is obtained by adding the pure three loop contribution from
  the asymptotic formula to the two--loop result (NNLO asympt.+ full
  NLO). The error band comes from the uncertainties in the low energy
  constants and is only shown for the best estimate. In the region above
  the $\mpi L = 2$ line, one is not safely in the $p$--regime and our
  results should not be trusted.
\label{fig:Rmm}}
\end{figure}

\begin{figure}[H]
\begin{center}
\includegraphics[width=12cm]{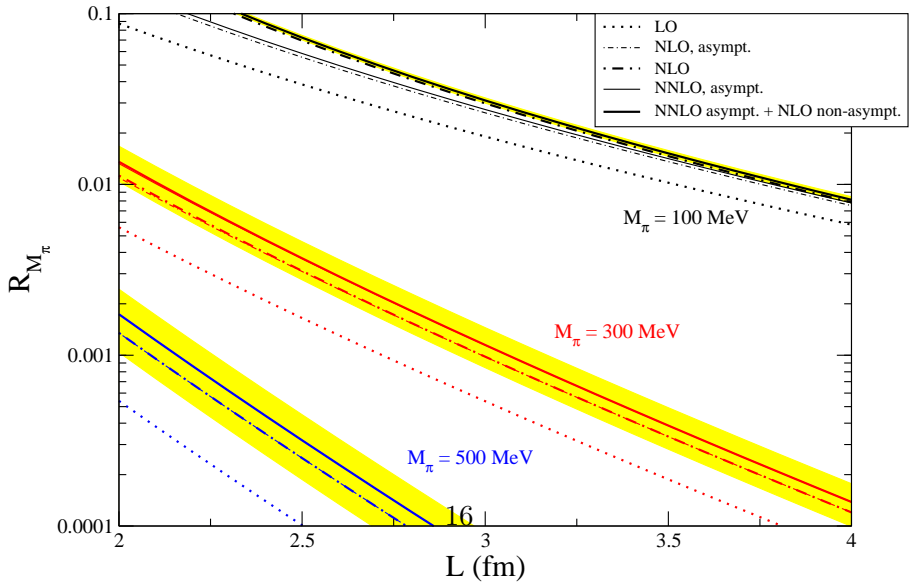}
\end{center}
\caption{$R_{\mpi}=\mpiL/\mpi-1$ 
  vs. $L$ for $\mpi=100,300,500$MeV. The rest of the legenda is as in
  fig.~\protect{\ref{fig:Rmm}}.
\label{fig:Rml}}
\end{figure}

\noindent
We plot our results for $R_{\mpi}$ in figs.~\ref{fig:Rmm},
and~\ref{fig:Rml} both for $L=2,3,4$fm as a function of $\mpi$ and for
$\mpi=100,300,500$MeV as a function of $L$.  We show the one--loop result
(LO) as well as the two--loop result (NLO). These shall be compared with
the resummed asymptotic formula with LO/NLO/NNLO input for the $\pi\pi$
scattering amplitude. Note that the one--loop result and the resummed
asymptotic formula to LO coincide. The best estimate for $R_{\mpi}$ is
finally obtained by adding to the asymptotic pure three--loop contribution
the two--loop result (NNLO asympt. + full NLO).
At NLO, the finite size effects contain low energy constants, see eg.
diagram $f)$ and $g)$ in fig.~\ref{fig:selfe}, leading to a non--negligible
error band.
\newpage
\addtolength{\topmargin}{-2cm}
\newpage
\addtolength{\topmargin}{2cm}
\begin{table}[t]
\small{
\begin{center}
\begin{tabular}{|llr|lll|}
\hline
\rule[-1mm]{0mm}{5mm}         
$R_{\mpi}$      &
               &
               &
$\mpi=140$ MeV &
$\mpi=200$ MeV &
$\mpi=250$ MeV\\
\rule[-2mm]{0mm}{5mm}         
$L=2$ fm       &
               &
               &
$\mpi L \simeq 1.4$ &
$\mpi L \simeq 2.0$ &
$\mpi L \simeq 2.5$ \\
\hline
               &
$I_p$          &
LO             &
0.0463         &
0.0199         &
0.0105         \\
               &
$I_p$          &
$\Delta$NLO    &
0.0291         &
0.0149         &
0.0088         \\
               &
$I_c$          &
$\Delta$NLO    &
0.0005         &
0.0001         &
0.0000         \\
               &
$\Sigma^{(2)}$ &
$\Delta$NLO    &
0.0206         &  
0.0038         &  
0.0011         \\ 
               &
$I_p$          &
$\Delta$NNLO   &
0.0076         &  
0.0053         &  
0.0035         \\ 
\hline
\hline
\multicolumn{3}{|l|}{\rule[-3mm]{0mm}{8mm}         
Total}
               &
0.1041(55)     &
0.0440(50)     &
0.0239(42)     \\
\hline
\end{tabular}
\end{center}}
\caption{$R_{\mpi}=\mpiL/\mpi-1$ for a $2$ fm volume and pion masses
  $\mpi=140$MeV, 200MeV, 250MeV. We show the relative numerical  impact of the
  LO, the pure NLO and the pure NNLO contribution for $R_{\mpi}$. In the
  second column we give the source of the effect. The fifth line,
  e.g. contains the contribution to $R_{\mpi}$ of the dispersive terms
  $I_c$ alone, once kinematical prefactors are added accordingly. The
  numerical results for the $I_p$ contributions, to LO, NLO and NNLO have
  already been given in~\protect\cite{Colangelo:2003hf,Colangelo:2005gd}.}
\label{3tab:num}
\end{table}

We take up a point which was already observed in \cite{Colangelo:2002hy},
namely the large contributions when going from LO to NLO in the asymptotic
formula (dotted to thin--dash--dotted). Compared to this gap, the additional
contributions from the full two--loop result (thick--dash--dotted) are very
small. 
The two--loop and the NLO result from the asymptotic formula only drift
away, when we go beyond the region where the $p$--regime can be safely
applied. In tab.~1 we wish to underline these statements with a numerical
example: we show the relative numerical impact of the LO, the pure NLO and
the pure NNLO contribution for $R_{\mpi}$. In the second column we give the
source of the effect. The fifth line, e.g. contains only the contribution
of the dispersive terms $I_c$ to $R_{\mpi}$, once the kinematical
prefactors are properly accounted for. We observe that $I_c$ is strongly
suppressed, irrespective of the value $\mpi L$. Consider the column with
$\mpi L\simeq 1.4$.  Although the bulk of the subleading effects is still
due to the asymptotic contributions, the terms of $\Sigma^{(2)}$ play a
significant role and can not be neglected. This behaviour was expected,
since with $\mpi L\simeq 1.4$ we might have already crossed the border of the
$p$--regime.  As we increase this parameter to $\mpi L \simeq 2$, the
asymptotic regime begins to set in. The contributions from the resummed
asymptotic formula are now dominating with respect to those from
$\Sigma^{(2)}$. The numerical results for $\mpi L\simeq 2.5$ confirm this
trend. 
The fact that even the additional NNLO asymptotic terms (a partial
three--loop result) are larger than the NLO non--asymptotic contributions
is in nice agreement with the analytical expectation found in
eq.~(\ref{eq:poly_mpiL}). However, we find that the suppression seen in the
numbers is stronger than what one could have expected on the basis of the
latter analytical argument. Finally, the discussion of the subleading
effects allows us to give a reliable estimate for the lower bound of $\mpi
L$ for the $p$--regime,
\begin{equation}
  \mpi L \gtrsim 2 \, :  \quad \mbox{lower bound for
    $p$--regime} 
  \fs 
\end{equation}

\section{Summary}

\noindent
\begin{itemize}
\item[\emph{i)}] We have evaluated the finite volume corrections for the pion
  mass to two--loops within the framework of chiral perturbation theory (ChPT)
  in the $p$--regime $(\mpi L\gg1, L> 2 \mbox{fm},\mpi<500 \mbox{MeV})$. 
\item[\emph{ii)}] We have compared the two--loop result with the resummed
  version of the asymptotic formula of L\"uscher. We have found that
  whenever the effects are calculated for masses and volumes such that
  $M_\pi L \gg1$, such that one is safely within the p--regime of ChPT, the
  contributions which are not included in the resummed asymptotic formula
  are very small. The result gives us confidence in the claim that the
  resummed asymptotic formula is a convenient and efficient way to reliably
  calculate finite volume effects for hadronic
  masses~\cite{Colangelo:2005gd}. 
\item[\emph{iii)}] The derivation of the asymptotic formula for decay
  constants~\cite{Colangelo:2004xr} follows closely the original one for
  the masses~\cite{Luscher:1985dn} -- the only new, subtle point, concerns
  the amplitude which appears in the integrand of the asymptotic formula,
  which has a pole in the integration region which first needs to be
  subtracted (see~\cite{Colangelo:2004xr} for details). This is, however, a
  technical point -- the physics of the finite volume corrections for
  masses and decay constants appears to be rather similar. In view of this
  we believe that the present results speak also in favour of the resummed
  asymptotic formula for decay constants \cite{Colangelo:2005gd}. A check
  of this claim could be obtained by evaluating the pion decay constant to
  two loops.
\item[\emph{iv)}] The two--loop calculation performed here allows us to
  better estimate the region of validity of ChPT in the $p$--regime. Our
  conclusion is that in the case of the pion mass it is necessary to have
  $\mpi L \gtrsim 2$. Again, this serves as a guideline also for decay
  constants and masses of other hadrons.
\end{itemize}

\subsection*{Acknowledgments}
We thank Stephan D\"urr for useful discussions and a careful reading of the
manuscript. This work has been supported by the Schweizerischer
Nationalfonds and partly by the EU ``Euridice'' program under code
HPRN-CT2002-00311. 

\appendix

\section{Finite volume integrals}
\label{app:1}

\noindent
In this appendix we give further details on the finite volume integrals
which occurred in the two--loop calculation. Throughout we applied
dimensional regularization, as is common in ChPT. Since the finite box
breaks Lorentz invariance, tensor simplifications have to be performed with
care. Consider e.g. the integral
\begin{equation}
  \hat p_\mu \hat p_\nu A_{\mu\nu} = 
  \int\frac{d^dq}{(2\pi)^d}\,
  \sum_{{\mb n} \in \mathbb{Z}^3}{}^{'} \,
  \frac{e^{\ri {\mb q\mb n}L}}{\mpi^2+q^2} 
  (\hat pq)^2 \co
\end{equation}
which arises when evaluating diagram $f)$ in fig.~\ref{fig:selfe}. Remember
$\hat{p}=(\ri\mpi,{\bf 0})$. The prime in the sum denotes that the term with
${\bf n}={\bf 0}$ is excluded. The ansatz
\begin{equation}
  \label{eq:ansatz}
  A_{\mu\nu} = \delta_{\mu\nu} A
\end{equation}
with $A$ a scalar integral which is valid in infinite volume, leads to an
incorrect result in finite volume. A direct calculation of
the integral with the help of eq.~(\ref{eq:schwinger}) yields
\begin{equation}
  \hat p_\mu \hat p_\nu A_{\mu\nu} = 
  -\frac{(\mpi^2)^{1+d/2}}{(2\pi\lapi)^{d/2}}
  \sum_{n=1}^{\infty}
  \frac{m_d(n)}{n^{d/4}}
  K_{d/2}(\lapi\sqrt{n})
  \fs
\end{equation}
In four dimensions the result agrees with the derivation outlined in the next
section, where one proceeds along the same lines as in sect.~\ref{sec:sig1}.

\subsection{Tadpole}

\noindent
In fig.~\ref{fig:selfe}, the finite volume corrections of the diagrams
$(a)$--$(c)$,$(f)$ and $(g)$ factorize into one--loop integrals which are
of the generic form 
\begin{equation}
  \int\frac{d^4q}{(2\pi)^4}\,
  \sum_{{\mb n} \in \mathbb{Z}^3}{}^{\prime} \,
  \frac{e^{\ri {\mb q\mb n}L}}{(\mpi^2+q^2)^k} 
  \cP(\hat pq;q^2)
  \co 
\end{equation}
with $k$ an integer positive number and $\cP(\hat pq,q^2)$ a polynomial of
$\hat pq$ and $q^2$. It suffices to discuss the case for $k=1$, since
$k=2,3,\ldots$ are obtained through appropriate derivatives with respect to
$\mpi^2$.  The only singularity of the integrand is the pole of the
propagator. Therefore, the contour integration analysis applied in
sect.~\ref{sec:sig1} yields the result
\begin{equation}
  \sum_{n=1}^{\infty} \frac{m(n)}{\sqrt{n}}
  \frac{\mpi^2}{8\pi^2\lapi}
  \int\limits_{-\infty}^{\infty}dy\, 
  e^{-\sqrt{n(1+y^2)}\lapi}\,
  \cP(\ri \mpi^2y;-\mpi^2) \fs
\end{equation}
In particular, in the text we have used the dimensionless function $\til
g_1(\lapi)$,
\begin{eqnarray}
  \til g_1(\lapi) &=& \frac{16\pi^2}{\mpi^2} g_1(\mpi^2,0,L) \co \nn
  g_k(\mpi^2,0,L) &=& 
  \sum_{{\mb n} \in \mathbb{Z}^3}{}^{\prime} \,
  \int
  \frac{d^d q}{(2\pi)^d} 
  \frac{e^{\ri \mb{q}\mb{n}L}}{(\mpi^2+q^2)^k} \co
\end{eqnarray}
evaluated at $d=4$, and
where $g_1(\mpi^2,0,L)$ was introduced a long time ago by
Gasser and Leutwyler \cite{Gasser:1986vb,Gasser:1987ah}, see also
eq.~(\ref{eq:g1dec}).

\subsection{Sunset}

\noindent
The only real two--loop diagram is the sunset graph
fig.~\ref{fig:selfe}$d)$, and we will comment on it in some detail. It is
convenient to split the finite volume integrals as in eq.(\ref{3tab:split})
only after the tensor simplifications. As alluded in the beginning of the
appendix, tensor simplifications in finite volume have to be performed with
care. Even though one can not rely on Lorentz invariance, the sunset tensor
integrals may still be reduced to the structures
\begin{equation}
  \{\cH,\cH_\mu,\cH_{\mu\nu}\} = \int d^dx e^{ipx}
  G(x)^2
  \Big[\{1,\ri\partial_\mu,-\partial_\mu\partial_\nu\} G(x) \Big]
  \co
\end{equation}
with $G(x)$ from eq.~(\ref{eq:Gx1}). A rather direct way to perform these
steps is to work in coordinate space and to make use of partial integrations,
i.e. 
\begin{eqnarray}
  \int d^dx e^{ipx}
  \partial_\mu G(x) \partial_\nu G(x) G(x) &=& \nn
  &&
  \hspace*{-1.8cm}
  -\frac{1}{2} \int d^dx e^{ipx} G(x)^2
  \bigg[ p_\mu i\partial_\nu G(x) 
  + \partial_\mu\partial_\nu G(x) \bigg] 
  \fs
\end{eqnarray}
Notice that the same identities in momentum space are derived with the help of 
translational invariance which is still respected in finite volume due to the
periodic boundary conditions. In the following, we will only elaborate on the
scalar integral $\cH$ in more detail. We expand the integral in terms of
number of finite volume propagators as motivated in
sect.~\ref{sect:two--loop},  
\begin{equation}
  \cH = \cH^{(0)} + 3 \cH^{(1)} + \cH^{(2)} \co
\end{equation}
where the first (second) addend corresponds to the pure (simple) gauge fields
contribution. For $\cH^{(0)}$ we refer to \cite{Gasser:1998qt}. Further,
\begin{equation}
  \label{eq:H1}
  \cH^{(1)} =   
  \sum_{n=1}^{\infty} m(n)
    \int \frac{d^4q}{(2\pi)^4}
  \frac{e^{i{q_1}{\sqrt{n}}L}}{\mpi^2+q^2}
  \bar J\left[(\hat{p}-q)^2\right]
  + g_1(\mpi^2,0,L) J(0) \co
\end{equation}
with
\begin{equation}
  J(k^2) =\int \frac{d^d\ell}{(2\pi)^d} 
  \frac{1}{[\mpi^2+\ell^2]}
  \frac{1}{[\mpi^2+(k-\ell)^2]}= \bar J(k^2) + J(0)
   \fs
\end{equation}
The first term of eq.~(\ref{eq:H1}) is finite, the second carries an
uv--divergence which is absorbed by a counterterm. This shows that although
finite volume effects do not generate new uv--divergences, they still appear at
intermediate steps of the calculation. It is a thorough check on our
calculation that these non--analytic divergences cancel. Finally,
\begin{equation}
  \cH^{(2)} = \sum_{{{\bf n},{\bf r}\in\mathbb{Z}^3\backslash{\bf 0}}
    \atop{{\bf n}\neq{\bf r}}}
  \int \frac{d^4q}{(2\pi)^4}\frac{d^4 k}{(2\pi)^4}
  \frac{e^{i{\bf q}{\bf n}L}}{[\mpi^2+q^2]}
  \frac{e^{i{\bf k}{\bf r}L}}{[\mpi^2+k^2]}
  \frac{1}{[\mpi^2+(\hat{p}-q-k)^2]} \fs
\end{equation}
In the text, we used its dimensionless version
\begin{equation}
  \{H;H_\mu;H_{\mu\nu}\} = 
  \sum_{{\mb{n},{\mb r}\in\mathbb{Z}^3\backslash{\mb 0}}
    \atop{{\mb n}\neq{\mb r}}}
  \int \frac{d^4\til q}{(2\pi)^4}\frac{d^4\til k}{(2\pi)^4}
  \frac{e^{i{\til\mb q}{\mb n}\lapi}}{[1+\til q^2]}
  \frac{e^{i{\til\mb k}{\mb r}\lapi}}{[1+\til k^2]}
  \frac{\{1;\til k_\mu;\til k_\mu \til k_\nu\}}
  {[1+(\til{p}-\til q-\til k)^2]} \co
\end{equation}
which are uv--finite and do not need to be renormalized. We only have to
find a convenient representation for the numerical analysis. We shall
restrict ourselves to the scalar integral in the following. Introducing a
Feynman parameter by combining the second and third denominator, we find
\begin{equation}
  \label{eq:Hscal}
  H = \int\frac{d^4\til q}{(2\pi)^4} \frac{d^4\til k}{(2\pi)^4} 
  \int\limits_0^1 dx
  \sum_{{\mb{n},{\mb r}\in\mathbb{Z}^3\backslash{\mb 0}}
    \atop{{\mb n}\neq{\mb r}}}
  \frac{1}{[1+\til q^2]} 
  \frac{e^{\ri\til\mb{q}(\mb{r}-\mb{n}x)\lapi+\ri\mb{n}\til\mb{k}\lapi}}
    {[1+(\til p-\til q)^2 x(1-x)+\til k^2]^2} \fs
\end{equation}
We use Schwinger's trick for both remaining denominators
\begin{equation}
  \label{eq:schwinger}
  \frac{1}{1+x^2} = \int\limits_0^\infty d\alpha \,
  e^{-\alpha(1+x^2)} \fs
\end{equation}
The integrals over $\til k$ and $\til q$ are then of the Gaussian type and
can be performed analytically. We are then
left with three integrations over a rather lengthy expression which shall not
be written down here. Despite their unhandy form, the integrations may
still safely be performed numerically. The accuracy of the determination of
eq.~(\ref{eq:Hscal}) is not limited by the integration routine, but by the
rather slow convergence of the sum in ${\bf n}$ and ${\bf r}$ for moderate
$\lapi$. Consequently, the evaluation of the sunset integrals going into
$\Delta$ is restricted to three significant digits. (The last digit given
for $\Sigma^{(2)}$ in tab.~1 is not significant.) 
Note that the uncertainty of the $H$--type integrals is not
a serious matter. Firstly, it could be lowered by brute force and secondly it
only plays a (minor) role, in case when the $p$--regime can not be safely
applied anymore. 

\section{Self--energy to second order: $\Sigma^{(2)}$}
\label{app:sigma2}

\noindent
In this section we ask ourselves whether the self--energy to second order can
be represented in a similar compact form as in the case of the self--energy
to first order in eq.~(\ref{eq:sig1}).  As will be discussed, it is indeed
possible to relate the self--energy to second order to a $3\pi$ scattering
amplitude in the forward scattering kinematics, as illustrated in
fig.~\ref{fig:skel}$b)$. This has already been observed by Schenk in a
related context \cite{Schenk:1993ru}. He investigated the dynamics of pions
in a cold heat bath of temperature $T$ and examined the effective mass of
the pion $M_\pi(T)$ within the framework of ChPT.  The close relation
between finite temperature field theory and finite volume effects becomes
apparent in the imaginary time formalism, where one treats the inverse
temperature as a finite extension in the imaginary time direction. The
expansion in terms of the number of finite volume propagators is then in
one--to--one correspondence with the expansion in terms of the number of
finite temperature propagators. In the following, Schenk's approach shall
be adapted to the finite volume scenario. We first establish the relation
between $\Sigma^{(2)}$ and the three--to--three particle scattering
amplitude and proceed with various remarks.

Consider the 6--point function in $d$ dimensions in the non--linear sigma
model parameterization eq.~(\ref{eq:sigma_param}) in the forward kinematics
\begin{eqnarray}
  \lefteqn{\sum\limits_{a,b=1}^{3}
  \int dx_1\ldots dx_5 e^{-\ri p(x_1-x_4) -\ri k(x_2-x_5) -\ri q(x_3-x_6)}
  \<0|\rm{T}\varphi^1_{x_1}\varphi^a_{x_2}\varphi^b_{x_3}\varphi^1_{x_4}
            \varphi^a_{x_5}\varphi^b_{x_6}|0\>}\nn
  &&= \frac{Z^3}{(\mpi^2+p^2)^2(\mpi^2+k^2)^2(\mpi^2+q^2)^2}
  T_{\pi\pi\pi}(p,k,q) \co
\end{eqnarray}
with $Z$ the wave function renormalization constant and $\varphi_x \equiv
\varphi(x)$. The amplitude
$T_{\pi\pi\pi}(p,k,q)$ contains a pole at $\hat{p}^2=-\mpi^2$ which needs to be
subtracted
\begin{equation}
  \label{eq:properT}
  T_{\pi\pi\pi}(p,k,q) = \hat{T}_{\pi\pi\pi}(p,k,q) + 
  \frac{R(p,k,q)}{\mpi^2+p^2} \fs
\end{equation}
\begin{figure}[t]
\begin{center}
\includegraphics[width=14cm]{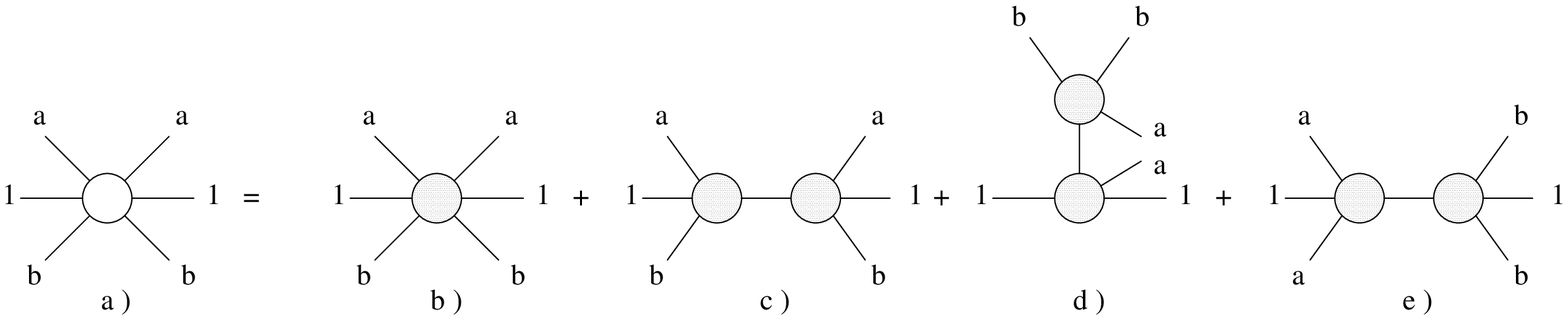} 
\end{center}
\caption{Decomposition of the $3\pi$--$3\pi$ amplitude in
  terms of 1particle irreducible parts. The characters $1$, $a$ and $b$ on the
  external legs denote isospin indices.\label{fig:6pi}}
\end{figure}  
The self--energy to second order can then be written in terms of the subtracted
amplitude $\hat{T}_{\pi\pi\pi}(\hat{p},k,q)$
\begin{equation}
  \label{eq:self2}
  \Sigma^{(2)} = \frac{1}{8} \sum\limits_{{\bf n},{\bf r}}^{} {}^\prime 
  \int\frac{d^4q}{(2\pi)^4}\frac{d^4k}{(2\pi)^4}
  e^{\ri {\bf qn}L+\ri {\bf kr}L}
  \frac{\hat{T}_{\pi\pi\pi}(\hat{p},k,q)}
  {(\mpi^2+q^2)(\mpi^2+k^2)}
   +\cO\left(\frac{1}{\fpi^6}\right) \; ,
\end{equation}
where the prime restricts the sum to
  integer vectors ${\bf n}$ and ${\bf r}$ obeying
  ${\bf n} \neq 0 \neq {\bf r}$, and for diagram $c)$ in fig.~\ref{fig:6pi}
  in addition ${\bf n} \neq {\bf r}$. 
  The latter restriction avoids double counting as this term is already
  accounted for in a simple gauge field of $\Sigma^{(1)}$.
\begin{itemize}
\item[\emph{i)}] The reason for the subtraction is easily accounted for. In
  fig.~\ref{fig:6pi} we decompose the $3\pi$--$3\pi$ scattering amplitude
  in terms of 1--particle irreducible parts. The subtraction removes the
  contribution from the diagram in fig.~\ref{fig:6pi}$e)$ which does not
  correspond to a 1--particle irreducible self--energy diagram, once the
  external legs are appropriately closed.

  A thorough discussion on the physical interpretation of the pole term can be
  found in \cite{Schenk:1993ru}.
\item[\emph{ii)}] Notice that eq.~(\ref{eq:properT}) defines a subtracted
  off--shell amplitude which depends on the regularization scheme as well
  as on the parameterization of the pion fields. At the order we are
  working, the regularization dependence is not an issue, since the
  subtracted amplitude $\hat{T}_{\pi\pi\pi}(\hat{p},k,q)$ is only needed at
  tree level. However, the dependence on the parameterization of the pion
  fields is of concern. While the momentum integrations in
  eq.~(\ref{eq:self2}) for the diagrams fig.~\ref{fig:6pi}$b)$ and
  \ref{fig:6pi}$d)$ put the momenta $k$ and $q$ on--shell and the ambiguity
  due to the parameterization therefore drops out, this does not happen in
  the case of diagram~\ref{fig:6pi}$c)$.

  Note that the same parameterization ambiguity already occurred in $I_c$
  in the dispersive analysis of $\Sigma^{(1)}$ (cf.
  eq.~(\ref{eq:off--shell}) where $q^2 \neq -\mpi^2$). In order to
  understand the close relation between these two terms, we first note that
  only diagram fig.~\ref{fig:selfe}$(d)$ contributed to $I_c$. Further, the
  simple gauge field of fig.~\ref{fig:selfe}$(d)$ with ${\bf n\neq 0}$ for
  one propagator can immediately be written as a contribution with two
  finite volume propagators: one periodifies a second propagator, however
  with the same ${\bf n\neq 0}$ as already for the first one. Since $\mpiL$
  does not depend on the parameterization of the pion fields, the
  ambiguities of the two terms have to cancel each other. To explicitly
  show this is however nontrivial.

  In summary, while our representations for $I_c$ in
  eq.~(\ref{eq:Ic1}), resp. eq.~(\ref{eq:Ic}) and for
  $\Sigma^{(2)}$ in eq.~(\ref{eq:self2}), resp. eq.~(\ref{eq:Sig2}) do
  depend on the off--shell dependence of the scattering amplitudes, the sum
  in eq.~(\ref{eq:sum1}) does not.
\item[\emph{iii)}] Even though the self--energy to second order can be
  expressed in a compact form, further simplifications (similar to those
  performed in sect.~\ref{sec:sig1} seem not to be straightforward.
  Instead, for the (numerical) integrations we had to discuss the various
  terms contributing to $\Sigma^{(2)}$ one by one. A general discussion of
  finite volume integrals has been provided in app.~\ref{app:1}.
  After treating the finite volume integrals, we end up with the basic
  functions given in eq.~(\ref{eq:Sig2}).
\end{itemize}


\newpage

\begin{thebibliography}{99}
\addcontentsline{toc}{chapter}{\hspace*{0.25cm} References}

\bibitem{Luscher:1985dn}
  M.~Luscher,
  %``Volume Dependence Of The Energy Spectrum In Massive 
  % Quantum Field Theories.
  %1. Stable Particle States,''
  Commun.\ Math.\ Phys.\  {\bf 104}, 177 (1986).
  %%CITATION = CMPHA,104,177;%%

\bibitem{Gasser:1986vb}
  J.~Gasser and H.~Leutwyler,
  %``Light Quarks At Low Temperatures,''
  Phys.\ Lett.\ B {\bf 184}, 83 (1987).
  %%CITATION = PHLTA,B184,83;%%

\bibitem{Gasser:1987ah}
  J.~Gasser and H.~Leutwyler,
  %``Thermodynamics Of Chiral Symmetry,''
  Phys.\ Lett.\ B {\bf 188}, 477 (1987).
  %%CITATION = PHLTA,B188,477;%%

\bibitem{Gasser:1987zq}
  J.~Gasser and H.~Leutwyler,
  %``Spontaneously Broken Symmetries: Eeffective Lagrangians At Finite
  %Volume,'' 
  Nucl.\ Phys.\ B {\bf 307}, 763 (1988).
  %%CITATION = NUPHA,B307,763;%%

\bibitem{Colangelo:2002hy}
  G.~Colangelo, S.~Durr and R.~Sommer,
  %``Finite size effects on M(pi) in QCD from chiral perturbation theory,''
  Nucl.\ Phys.\ Proc.\ Suppl.\  {\bf 119}, 254 (2003)
  [hep-lat/0209110].
  %%CITATION = HEP-LAT 0209110;%%

\bibitem{Colangelo:2003hf}
  G.~Colangelo and S.~Durr,
  %``The pion mass in finite volume,''
  Eur.\ Phys.\ J.\ C {\bf 33}, 543 (2004)
  [hep-lat/0311023].
  %%CITATION = HEP-LAT 0311023;%%

\bibitem{Gasser:1983yg}
  J.~Gasser and H.~Leutwyler,
  %``Chiral Perturbation Theory To One Loop,''
  Annals Phys.\  {\bf 158}, 142 (1984).
  %%CITATION = APNYA,158,142;%%

\bibitem{Bijnens:1995yn}
  J.~Bijnens, G.~Colangelo, G.~Ecker, J.~Gasser and M.~E.~Sainio,
  %``Elastic $\pi\pi$ scattering to two loops,''
  Phys.\ Lett.\ B {\bf 374}, 210 (1996)
  [hep-ph/9511397].
  %%CITATION = HEP-PH 9511397;%%

\bibitem{Bijnens:1997vq}
  J.~Bijnens, G.~Colangelo, G.~Ecker, J.~Gasser and M.~E.~Sainio,
  %``Pion pion scattering at low energy,''
  Nucl.\ Phys.\ B {\bf 508}, 263 (1997)
  [Erratum-ibid.\ B {\bf 517}, 639 (1998)]
  [hep-ph/9707291].
  %%CITATION = HEP-PH 9707291;%%

\bibitem{Colangelo:2001df}
  G.~Colangelo, J.~Gasser and H.~Leutwyler,
  %``pi pi scattering,''
  Nucl.\ Phys.\ B {\bf 603}, 125 (2001)
  [hep-ph/0103088].
  %%CITATION = HEP-PH 0103088;%%

\bibitem{Colangelo:2004sc}
  G.~Colangelo,
  %``Finite volume effects in chiral perturbation theory,''
  Nucl.\ Phys.\ Proc.\ Suppl.\  {\bf 140}, 120 (2005)
  [hep-lat/0409111].
  %%CITATION = HEP-LAT 0409111;%%

\bibitem{Colangelo:2005gd}
  G.~Colangelo, S.~Durr and C.~Haefeli,
  %``Finite volume effects for meson masses and decay constants,''
  Nucl.\ Phys.\ B {\bf 721} (2005) 136
  [arXiv:hep-lat/0503014].
  %%CITATION = HEP-LAT 0503014;%%

\bibitem{Haefeli:2005px}
  C.~Haefeli,
  %``The pion mass in finite volume to two loops,''
  [hep-lat/0509078].
  %%CITATION = HEP-LAT 0509078;%%

\bibitem{Colangelo:2005cg}
  G.~Colangelo, A.~Fuhrer and C.~Haefeli,
  %``The pion and proton mass in finite volume,''
  [hep-lat/0512002].
  %%CITATION = HEP-LAT 0512002;%%

\bibitem{Sharpe:1992ft}
  S.~R.~Sharpe,
  %``Quenched chiral logarithms,''
  Phys.\ Rev.\ D {\bf 46}, 3146 (1992)
  [hep-lat/9205020].
  %%CITATION = HEP-LAT 9205020;%%

\bibitem{Becirevic:2003wk}
  D.~Becirevic and G.~Villadoro,
  %``Impact of the finite volume effects on the chiral behavior of f(K) and
  %B(K),''
  Phys.\ Rev.\ D {\bf 69}, 054010 (2004)
  [hep-lat/0311028].
  %%CITATION = HEP-LAT 0311028;%%

\bibitem{AliKhan:2003cu}
  A.~Ali Khan {\it et al.}  [QCDSF-UKQCD Collaboration],
  %``The nucleon mass in N(f) = 2 lattice QCD: Finite size effects from  chiral
  %perturbation theory,''
  Nucl.\ Phys.\ B {\bf 689}, 175 (2004)
  [hep-lat/0312030].
  %%CITATION = HEP-LAT 0312030;%%

\bibitem{Arndt:2004bg}
  D.~Arndt and C.~J.~D.~Lin,
  %``Heavy meson chiral perturbation theory in finite volume,''
  Phys.\ Rev.\ D {\bf 70}, 014503 (2004)
  [hep-lat/0403012].
  %%CITATION = HEP-LAT 0403012;%%

\bibitem{Beane:2004tw}
  S.~R.~Beane,
  %``Nucleon masses and magnetic moments in a finite volume,''
  Phys.\ Rev.\ D {\bf 70}, 034507 (2004)
  [hep-lat/0403015].
  %%CITATION = HEP-LAT 0403015;%%

\bibitem{Beane:2004rf}
  S.~R.~Beane and M.~J.~Savage,
  %``Baryon axial charge in a finite volume,''
  Phys.\ Rev.\ D {\bf 70}, 074029 (2004)
  [hep-ph/0404131].
  %%CITATION = HEP-PH 0404131;%%


\bibitem{Bedaque:2006yi}
  P.~F.~Bedaque, I.~Sato and A.~Walker-Loud,
  %``Finite volume corrections to pi pi scattering,''
  [hep-lat/0601033].
  %%CITATION = HEP-LAT 0601033;%%
  %%Cited 0 times in SPIRES-HEP

\bibitem{Bijnens:2005ne}
  J.~Bijnens, N.~Danielsson, K.~Ghorbani and T.~Lahde,
  %``Two loop partially quenched and finite volume chiral perturbation theory
  %results,''
  [hep-lat/0509042].
  %%CITATION = HEP-LAT 0509042;%%

\bibitem{Schenk:1993ru}
  A.~Schenk,
  %``Pion propagation at finite temperature,''
  Phys.\ Rev.\ D {\bf 47}, 5138 (1993).
  %%CITATION = PHRVA,D47,5138;%%

\bibitem{Toublan:1997rr}
  D.~Toublan,
  %``Pion dynamics at finite temperature,''
  Phys.\ Rev.\ D {\bf 56}, 5629 (1997)
  [hep-ph/9706273].
  %%CITATION = HEP-PH 9706273;%%

\bibitem{Luscher:1983rk}
  M.~Luscher,
  %``On A Relation Between Finite Size Effects And Elastic Scattering
  %Processes,''
  DESY 83/116
%\href{http://www.slac.stanford.edu/spires/find/hep/www?r=desy\%2F83\%2F116}{SPIRES entry}
  {\it Lecture given at Cargese Summer Inst., Cargese, France, Sep 1-15, 1983}

\bibitem{Koma}
  Y.~Koma and M.~Koma,
  %``On the finite size mass shift formula for stable particles,''
  Nucl.\ Phys.\ B {\bf 713}, 575 (2005)
  [hep-lat/0406034].
  %%CITATION = HEP-LAT 0406034;%%
  Y.~Koma and M.~Koma,
  %``More on the finite size mass shift formula for stable particles,''
  [hep-lat/0504009].
  %%CITATION = HEP-LAT 0504009;%%

\bibitem{Braun:2004yk}
  J.~Braun, B.~Klein and H.~J.~Pirner,
  %``Volume dependence of the pion mass in the quark-meson model,''
  Phys.\ Rev.\ D {\bf 71}, 014032 (2005)
  [hep-ph/0408116].
  %%CITATION = HEP-PH 0408116;%%

\bibitem{Borasoy:2005nz}
  B.~Borasoy, G.~M.~von Hippel, H.~Krebs and R.~Lewis,
  %``Automated methods in chiral perturbation theory on the lattice,''
  [hep-lat/0509007].
  %%CITATION = HEP-LAT 0509007;%%

\bibitem{Bijnens:1999sh}
  J.~Bijnens, G.~Colangelo and G.~Ecker,
  %``The mesonic chiral Lagrangian of order p**6,''
  JHEP {\bf 9902}, 020 (1999)
  [hep-ph/9902437].
  %%CITATION = HEP-PH 9902437;%%

\bibitem{Bijnens:1999hw}
  J.~Bijnens, G.~Colangelo and G.~Ecker,
  %``Renormalization of chiral perturbation theory to order p**6,''
  Annals Phys.\  {\bf 280}, 100 (2000)
  [hep-ph/9907333].
  %%CITATION = HEP-PH 9907333;%%

\bibitem{Gasser:1998qt}
  J.~Gasser and M.~E.~Sainio,
  %``Two-loop integrals in chiral perturbation theory,''
  Eur.\ Phys.\ J.\ C {\bf 6}, 297 (1999)
  [hep-ph/9803251].
  %%CITATION = HEP-PH 9803251;%%

\bibitem{Hasenfratz:1989pk}
  P.~Hasenfratz and H.~Leutwyler,
  %``Goldstone Boson Related Finite Size Effects In Field Theory And Critical
  %Phenomena With O(N) Symmetry,''
  Nucl.\ Phys.\ B {\bf 343}, 241 (1990).
  %%CITATION = NUPHA,B343,241;%%

\bibitem{Burgi:1996qi}
  U.~Burgi,
  %``Pion polarizabilities and charged pion pair production to two loops,''
  Nucl.\ Phys.\ B {\bf 479}, 392 (1996)
  [hep-ph/9602429].
  %%CITATION = HEP-PH 9602429;%%

\bibitem{Colangelo:2004xr}
  G.~Colangelo and C.~Haefeli,
  %``An asymptotic formula for the pion decay constant in a large volume,''
  Phys.\ Lett.\ B {\bf 590}, 258 (2004)
  [hep-lat/0403025].
  %%CITATION = HEP-LAT 0403025;%%

\end{thebibliography}
\end{document}